\documentclass[a4paper,11pt]{article}
\usepackage{graphicx,rotating,hyperref,slashed,amsmath,xcolor,amssymb,amsfonts,colortbl,cite,subcaption}
\usepackage[totalheight = 23cm, totalwidth = 17cm]{geometry}
\usepackage{physics}

\definecolor{bluc}{cmyk}{1,1,0,0.1}
\definecolor{rossoCP3}{cmyk}{0,.88,.77,.40}
\definecolor{rosso}{cmyk}{0,1,1,0.4}
\definecolor{rossos}{cmyk}{0,1,1,0.55}
\definecolor{rossoc}{cmyk}{0,1,1,0.2}
\definecolor{verdes}{cmyk}{0.92,0,0.59,0.4}

\hypersetup{colorlinks, bookmarksopen, bookmarksnumbered,
citecolor=verdes, linkcolor=bluc, pdfstartview=FitH, urlcolor=rossos}

\newcommand{\mpl}{m_{\rm Pl}}
\newcommand{\fnl}{f_{\rm NL}}

\begin{document}

\begin{titlepage}

\rightline{\footnotesize{APCTP-Pre2023-003, YITP-23-23}}

\begin{center}

\vskip 3em

{\LARGE \bf 
New shape of parity-violating graviton non-Gaussianity
}

\vskip 3em

{\large
Jinn-Ouk Gong$^{a,b}$,
Maria Mylova$^{a}$ 
and
Misao Sasaki$^{c,d,e}$
}

\vskip 0.5cm

{\it
$^{a}$Department of Science Education,  Ewha Womans University, Seoul 03760, Korea
\\
$^{b}$Asia Pacific Center for Theoretical Physics, Pohang 37673, Korea
\\
$^{c}$Kavli Institute for the Physics and Mathematics of the Universe (WPI)
\\
The University of Tokyo, Kashiwa, Chiba 277-8583, Japan
\\
$^{d}$Center for Gravitational Physics and Quantum Information
\\
Yukawa Institute for Theoretical Physics, Kyoto University, Kyoto 606-8502, Japan
\\
$^{e}$Leung Center for Cosmology and Particle Astrophysics
\\ 
National Taiwan University, Taipei 10617, Taiwan
}

\end{center}

\vskip 1.2cm

\begin{abstract}

We show that the general vacuum states that respect the de Sitter symmetry, known as the $\alpha$-vacua, can introduce non-vanishing parity-violating tensor non-Gaussianities. This is due to the mixing by the Bogoliubov transformation of the positive and negative frequency modes of the Bunch-Davies vacuum. We calculate explicitly the bispectra of tensor perturbations and show that the amplitude can be exponentially enhanced for certain choices of the squeezing parameter $\alpha$ and the phase $\phi$ of the $\alpha$-vacua. We find a new shape for the parity-violating tensor bispectrum which peaks in the flattened configuration.

\end{abstract}

\end{titlepage}

\newpage

\section{Introduction}
\label{sec:intro}

Parity (P) violation in the weak interaction sector is of crucial importance to our understanding of the standard model of particle physics~\cite{Lee:1956qn,Wu:1957my}. Probably, the origin of parity violation lies in the fundamental physics whose low-energy effective theory is the standard model. There are tantalizing hints. For example, the baryon asymmetry that guarantees the very existence of the observable material world demands the violation of charge conjugation (C) as well as CP~\cite{Sakharov:1967dj}. Likewise any extension beyond the standard model may well incorporate non-standard interactions that give rise to parity violation. Furthermore, if exists, such a parity-violating interaction can leave observable signatures in the universe. For example, axion can couple to the electromagnetic field tensor via a Chern-Simons coupling, leading to the rotation of the polarization angle of the cosmic microwave background (CMB), known as cosmic birefringence~\cite{Ni:1977zz,Carroll:1989vb}. Indeed, recent analyses suggest a non-zero rotation angle at 3.6 $\sigma$ level~\cite{Minami:2020odp,Eskilt:2022cff,Diego-Palazuelos:2022cnh}, indicating parity violation in the matter sector beyond weak interaction. Then, we can naturally ask: Why not in the gravity sector?

One interesting aspect of the stochastic gravitational wave background is that it could be parity-violating, provided that general relativity can be extended to include terms that violate P and time-reversal invariance~\cite{Jackiw:2003pm}. This is permitted in the context of effective field theory to include derivative operators that breaks discrete spatial symmetries~\cite{Cannone:2015rra}. This can allow for a difference in the intensity of the left- and right-circular polarizations of gravitational waves, resulting in a preferred macroscopic orientation in the universe. It has been shown that if such an asymmetry existed at early times, it could have left observable traces in the CMB by producing non-vanishing TB and EB correlations~\cite{Lue:1998mq}. Additionally, we expect that higher-order interactions could contribute to parity violation in the CMB bispectra and higher-order correlation functions. Upcoming experiments will measure the B-mode polarization anisotropies with an expected precision $r \sim 10^{-3}$~\cite{Verde:2005ff,Amblard:2006ef,CMBPolStudyTeam:2008rgp,Hazumi:2019lys}. Parity-violating gravitational waves with sufficiently large amplitude could be measured with interferometers~\cite{Bartolo:2018qqn,Domcke:2019zls,Orlando:2020oko,Omiya:2023rhj}. A tantalizing possibility comes from more recent suggestions that there may be evidence of parity violation in the large-scale structure~\cite{Masui:2017fzw,Biagetti:2020lpx,Philcox:2022hkh}. Confirmation of any of these signals will give us a window into the primordial origin of our universe.

Parity-violating contributions to gravity appear naturally when we extend Einstein gravity to include higher-order curvature invariants~\cite{Weinberg:2008hq}. The leading-order parity-violating term is given by the gravitational four-dimensional Chern-Simons term $f \, W \widetilde{W}$~\cite{Lue:1998mq}, where $W_{\mu\nu\rho\sigma}$ is the Weyl tensor and $\widetilde{W}^{\mu\nu\rho\sigma} \equiv \epsilon^{\mu\nu\kappa\lambda} W_{\kappa\lambda}{}^{\rho\sigma}$. This term is topological and therefore contributes to parity violation only if the coupling is a generic function of some scalar field, $f=f(\phi)$. Unfortunately, theoretical predictions tend to suffer due to the Chern-Simons instability~\cite{Alexander:2004wk,Dyda:2012rj}, leading to negligibly small parity-violating two-point statistics, unless non-standard inflationary scenarios are considered~\cite{Satoh:2007gn,Satoh:2008ck,Satoh:2010ep,Sorbo:2011rz,Cai:2016ihp,Maleknejad:2012wqk,Mylova:2019jrj}.
Still, even in the most optimistic case, we do not expect to obtain sensible constraints for parity violation from the two-point statistics of the CMB as, even at low significance, this would require a maximally parity-violating signal with a fairly large tensor-to-scalar ratio. For this reason, attention has been focused on the three-point statistics of the CMB. The bispectrum for the gravitational Chern-Simons term was previously computed in~\cite{Bartolo:2017szm,Bartolo:2018elp}. There, it was found that only the scalar-scalar-tensor bispectrum is not suppressed, but it is still subject to the constraints from the Chern-Simons instability.

Another interesting possibility is to consider the case where a coupling to the inflaton is not necessary. The leading parity-violating term we are allowed to write is the six-dimensional parity-violating cubic Weyl term, $\widetilde{W} W^2$. It was previously shown that there is no parity-violating tensor bispectrum for the Bunch-Davies (BD) vacuum in the exact de Sitter space~\cite{Maldacena:2011nz}, by using the fact that the isometries of de Sitter space lead to conformally invariant correlation functions. This was confirmed in~\cite{Soda:2011am} by direct calculation of the tensor bispectra. These considerations were further extended by explicitly breaking the assumption of perfect de Sitter and found that there exists parity-violating non-Gaussianity during slow-roll inflation, with an enhancement proportional to the slow-roll parameter. From this, it follows that parity-violating non-Gaussianity is slow-roll suppressed unless one considers non-standard models of inflation. Another possibility is to generalize the generic coupling of the parity-violating cubic Weyl term to a dilaton-like coupling, which was examined in~\cite{Shiraishi:2011st}. Similar considerations apply to the ghost-free parity-violating theory of gravity proposed in~\cite{Crisostomi:2017ugk} and further examined in~\cite{Qiao:2019hkz,Bartolo:2020gsh}. The results in~\cite{Maldacena:2011nz,Soda:2011am} were more recently confirmed in~\cite{Bordin:2020eui,Cabass:2021fnw}, where it was shown that the parity-violating cubic Weyl term only contributes a phase to the graviton bispectrum. Additionally, in~\cite{Cabass:2021fnw} the bootstrap approach was employed to derive the most general model-independent graviton bispectra assuming scale invariance and BD initial condition. There, it was found that there can be at most three parity-odd graviton bispectra of the equilateral and squeezed shapes.

In this work, we entertain an alternative possibility. We study the parity-violating tensor non-Gaussianity in exact de Sitter space, but choose to loosen the assumption that the initial state is in the BD vacuum. Deviations from the typical adiabatic vacuum can be described by the $\alpha$-vacua which are known to be de Sitter invariant~\cite{Allen:1985ux,Higuchi:2011vw}, 
and have been largely used to address the effects of fluctuations of trans-Planckian origin (see for example \cite{Danielsson:2002kx,Danielsson:2002qh} and references therein). It was recently shown that depending on the parameters of the $\alpha$-vacuum, there can be large tensor non-Gaussianity in the squeezed and flattened limit of the bispectrum in Einstein gravity~\cite{Kanno:2022mkx,Ghosh:2023agt}.
Here, we extend these considerations to address the parity-violating tensor non-Gaussianity in the $\alpha$-vacuum. In particular, we focus on the parity-violating cubic Weyl term, being the dominant source of chirality in exact de Sitter space. Unlike previous works on the graviton correlations which were limited only to the typical BD vacuum, here we find the following key results:
\begin{itemize}
    \item Non-vanishing parity-violating tensor non-Gaussianity in exact de Sitter space. This is because the bispectra acquire an imaginary part that results from the  mixing of the positive and negative frequency modes by the Bogoliubov transformation. 
    \item A new shape for the parity-violating non-Gaussianity peaking in the flattened configuration, which is a natural consequence of interference among graviton modes inside the Hubble horizon. 
    \item The potential enhancement of the bispectra amplitudes, assuming sufficient deviations from the BD vacuum. 
\end{itemize}
We would like to emphasize that even a small deviation from the BD vacuum is sufficient to obtain non-vanishing tensor bispectra. This makes our results all the more important, because loosening the assumption of the perfect BD initial condition can allow much richer phenomenologies in the statistics of the CMB.

The work is organized as follows: The key ingredients for computing the tensor correlation functions are given in Section~\ref{sec:parity}. There, we show that the Bogoliubov mixing of the positive and negative frequency modes leads to different results from those with the BD initial condition. The physical contributions to the tensor bispectrum are computed in Section~\ref{sec:bi}, where we find non-vanishing parity-violating tensor non-Gaussianity in exact de Sitter background, with the dominant contributions coming from the mixed polarizations in the flattened configuration. We show the bispectrum amplitude can be exponentially enhanced for large squeezing parameter $\alpha$ in the limit where the phase approaches the value $\phi \rightarrow \pi$. Finally, we conclude in Section~\ref{sec:conc}.

\section{Parity-violation in $\alpha$-vacuum}
\label{sec:parity}

\subsection{Tensor perturbations}

To begin with, we first define tensor perturbations $h_{ij}$ in the flat Friedmann universe:
\begin{equation}
ds^2 = a^2(\eta) \big[ -d\eta^2 + ( \delta_{ij} + h_{ij} ) dx^idx^j \big] \, .
\end{equation}
Here, $d\eta \equiv dt/a$ is the conformal time with $a$ being the scale factor, and $h_{ij}$ is transverse and traceless: $h^i{}_{j,i} = 0$ and $h^i{}_i = 0$, with $i$ and $j$ being spatial indices which are raised and lowered by $\delta_{ij}$. The transverse and traceless conditions leave two physical degrees of freedom for $h_{ij}$, which we identify as the two polarizations of tensor perturbations, or gravitational waves. If we expand the Einstein-Hilbert action up to second-order in $h_{ij}$, we find
\begin{equation}
\label{eq:S2}
S_\text{EH} = \frac{\mpl^2}{8} \int d\eta d^3x \Big[  \big(h_{ij}'\big)^2 - \big(h_{ij,k}\big)^2 \Big] \, ,
\end{equation}
where a prime denotes a derivative with respect to the conformal time and $\mpl \equiv 1/\sqrt{8\pi G}$ is the reduced Planck mass. Since the quadratic action \eqref{eq:S2} is essentially the same as that of a harmonic oscillator, we can quantize $h_{ij}$ by introducing creation and annihilation operators. For this purpose, it is convenient to expand $h_{ij}$ in terms of its Fourier modes and to introduce the polarization tensor $e_{ij}^{(s)}(\pmb{k})$, with $s=L, R$ denoting the left- and right-circular polarizations of gravitational waves. That is, $e_{ij}^{(s)}(\pmb{k})$ satisfies the following properties:
\begin{align}
\label{eq:indices}
e_{ij}^{(s)} & = e_{ji}^{(s)} \, ,
\\
\label{eq:traceless}
\delta^{ij}e_{ij}^{(s)} & = 0 \, ,
\\
\label{eq:transverse}
k^ie_{ij}^{(s)} & = 0 \, ,
\\
\label{eq:2pol}
e_{ij}^{(s)} {e_{ij}^{(s')}}^* & = \delta_{ss'} \, ,
\\
\label{eq:helicity}
\frac{k_l}{k}\epsilon^{ilk}e_{jk}^{(s)} & = -ise^i{}_j^{(s)} \, ,
\\
\label{eq:realness}
e_{ij}^{(s)}(-\pmb{k}) & = {e_{ij}^{(s)}}^*(\pmb{k}) \, ,
\end{align}
which represent respectively the following: \eqref{eq:indices} means the symmetry between the spatial indices, \eqref{eq:traceless} tracelessness, \eqref{eq:transverse} transverseness, \eqref{eq:2pol} the existence of two independent polarizations, \eqref{eq:helicity} helicity states with $s=-1$ for left- and $s=+1$ for right-circular polarizations, and \eqref{eq:realness} the realness of $h_{ij}$. Then, we can expand $h_{ij}$ as
\begin{equation}
\label{eq:hij}
h_{ij}(\eta,\pmb{x}) =\frac{2}{m_{\rm Pl}} \int \frac{d^3k}{(2\pi)^{3/2}}  \frac{e^{i\pmb{k}\cdot\pmb{x}}}{\sqrt{2k}} 
\Big[
u_k(\eta) e_{ij}^{(s)}(\pmb{k}) a_{(s)}(\pmb{k}) + u_k^*(\eta) {e_{ij}^{(s)}}^*(-\pmb{k}) a_{(s)}^\dag(-\pmb{k})
\Big]
\, ,
\end{equation}
where the creation and annihilation operators obey the following commutation relations:
\begin{equation}
\Big[ a_{(s)}(\pmb{k}), a_{(s')}^\dag(\pmb{k}') \Big] = \delta_{ss'}\delta^{(3)}(\pmb{k}-\pmb{k}') \, ,
\end{equation}
otherwise zero. Then, the following normalization condition for the mode function $u_k(\eta)$ is imposed:
\begin{equation}
u_k^*u_k' - u_k{u_k^*}' = -\frac{2ik}{a^2} \, .
\end{equation}

\subsection{Cubic Weyl action}

As explained, we want to compute the contribution from the parity-violating cubic Weyl term~\cite{Maldacena:2011nz}:
\begin{equation}
\label{eq:cubicWeyl}
S_{\widetilde{W}W^2} = -\frac{b}{m_{\rm Pl}^2} \int d\eta d^3x \sqrt{-g} \epsilon^{\mu\nu\rho\sigma}
W_{\mu\nu\alpha\beta} W^{\alpha\beta\gamma\delta} W_{\gamma\delta\rho\sigma}
\, ,
\end{equation}
where $b$ is an arbitrary dimensionless coupling constant. In what follows, we adapt the methods in~\cite{Soda:2011am} to calculate the interaction Hamiltonian. Using the pseudo self-duality of the Weyl tensor, we can define the helicity eigenstates:
\begin{equation}
\label{eq:helicity-modes}
\gamma^{\pm}_{ij} \equiv \frac{1}{2} \big( \gamma^\prime_{ij}  \mp i \epsilon_{jkl} \gamma_{ik,l} \big) \, ,
\end{equation}
where $\gamma_{ij}$ is defined in terms of $h_{ij}$ as
\begin{equation}
\label{eq:gamma'}
\gamma_{ij}' \equiv ah_{ij}' \, .
\end{equation}
From this, we can write \eqref{eq:cubicWeyl} in terms of $\gamma_{ij}^\pm$ as
\begin{equation}
\label{eq:cubicWeyl2}
S_{\widetilde{W}W^2} = 8i\frac{b}{m_{\rm Pl}^2} \int d\eta d^3x a^{-5} 
\Big( 
{\gamma_{ij}^+}'{\gamma_{jk}^+}'{\gamma_{ki}^+}' - {\gamma_{ij}^-}'{\gamma_{jk}^-}'{\gamma_{ki}^-}'
\Big) 
\, .
\end{equation}
Note that only time derivatives of $\gamma_{ij}^\pm$ appear in $S_{\widetilde{W}W^2}$. From this, we can write immediately the corresponding cubic interaction Hamiltonian as
\begin{align}
\label{eq:H3}
H_{\widetilde{W}W^2}
& = 
-8i\frac{b}{m_{\rm Pl}^2a^5} \int \frac{d^3k_1 d^3k_2 d^3k_3}{(2\pi)^6} \delta^{(3)}(\pmb{k}_1+\pmb{k}_2+\pmb{k}_3)
\nonumber\\
& \hspace{5em}
\times
\Big[ 
{\gamma_{ij}^+}'(\eta,\pmb{k}_1){\gamma_{jk}^+}'(\eta,\pmb{k}_2){\gamma_{ki}^+}'(\eta,\pmb{k}_3) 
- {\gamma_{ij}^-}'(\eta,\pmb{k}_1){\gamma_{jk}^-}'(\eta,\pmb{k}_2){\gamma_{ki}^-}'(\eta,\pmb{k}_3)
\Big]
\, . 
\end{align}
Once we decompose \eqref{eq:H3} in terms of \eqref{eq:hij}, it is straightforward to  calculate the physical bispectra of the tensor perturbations. For this purpose, we usually need to find the mode function $u_k(\eta)$, which amounts to specifying the vacuum state annihilated by $a_{(s)}(\pmb{k})$.

\subsection{Mode expansion in $\alpha$-vacuum}

A vacuum state satisfies, for all $\pmb{k}$,
\begin{equation}
a_{(s)}(\pmb{k})|0_{\pmb{k}}\rangle = 0 \, .
\end{equation}
Demanding that the vacuum state be annihilated only by the positive frequency mode gives the BD vacuum. The corresponding mode function in an exact de Sitter background, where $a = -1/(H\eta)$ with $\eta < 0$ and the Hubble parameter $H$ being constant, is given by
\begin{equation}
u_k(\eta) = \frac{H}{k} (1+ik\eta) e^{-ik\eta} \, .
\end{equation}
But the BD vacuum is not the only one that respects the isometries of de Sitter space. There are infinitely many vacua that can be obtained by linear combinations of the BD mode function and operators [see \eqref{eq:alpha-mode} and \eqref{eq:alpha-annihilation}], called the $\alpha$-vacua. Therefore, we may consider another expansion of $h_{ij}$ as
\begin{equation}
\label{eq:hij-alpha}
h_{ij}(\eta,\pmb{x}) =\frac{2}{m_{\rm Pl}} \int \frac{d^3k}{(2\pi)^{3/2}} \frac{e^{i\pmb{k}\cdot\pmb{x}}}{\sqrt{2k}} 
\Big[
v_k(\eta) e_{ij}^{(s)}(\pmb{k}) c_{(s)}(\pmb{k}) + v_k^*(\eta) {e_{ij}^{(s)}}^*(-\pmb{k}) c_{(s)}^\dag(-\pmb{k})
\Big]
\, ,
\end{equation}
where
\begin{equation}
\Big[ c_{(s)}(\pmb{k}), c_{(s')}^\dag(\pmb{k}') \Big] = \delta_{ss'}\delta^{(3)}(\pmb{k}-\pmb{k}') \, ,
\end{equation}
and the operator $c_{(s)}(\pmb{k})$ annihilates the $\alpha$-vacuum:
\begin{equation}
c_{(s)}(\pmb{k})|0_{\pmb{k}}\rangle_\alpha = 0 \, .
\end{equation}
The mode function $v_k$ and the creation and annihilation operators of the $\alpha$-vacuum, $c_{(s)}^\dag(\pmb{k})$ and $c_{(s)}(\pmb{k})$, are related to those of the BD vacuum by the so-called Bogoliubov transformation:
\begin{align}
\label{eq:alpha-mode}
v_k(\eta) & = \cosh\alpha u_k(\eta) + e^{i\phi}\sinh\alpha u_k^*(\eta) \, ,
\\
\label{eq:alpha-annihilation}
c_{(s)}(\pmb{k}) & = \cosh\alpha a_{(s)}(\pmb{k}) - e^{-i\phi}\sinh\alpha a_{(s)}^\dag(-\pmb{k}) \, ,
\end{align}
where we assume $0\leq\alpha\leq\infty$ and $0\leq\phi\leq2\pi$ without losing generality. Now, we are ready to write $\gamma_{ij}^\pm$ in terms of $v_k$, $c_{(s)}^\dag(\pmb{k})$ and $c_{(s)}(\pmb{k})$. Substituting \eqref{eq:hij-alpha} into \eqref{eq:gamma'}, we can write \eqref{eq:helicity-modes} as
\begin{align}
{\gamma_{ij}^\pm}'(\eta,\pmb{x})
=
\int \frac{d^3k}{(2\pi)^{3/2}} \frac{e^{i\pmb{k}\cdot\pmb{x}}}{\sqrt{2k}\mpl} \sum_s
\Big[ e_{ij}^{(s)}(\pmb{k}) (\partial_\eta \mp isk) av_k'(\eta) c_{(s)}(\pmb{k})
+ {e_{ij}^{(s)}}^*(-\pmb{k}) (\partial_\eta \mp isk) a{v_k^*}'(\eta) c_{(s)}^\dag(-\pmb{k}) \Big]
\, .
\end{align}
Then, using the perfect de Sitter background $a = -1/(H\eta)$, we can find the solutions of $\gamma_{ij}^\pm$ as
\begin{align}
\label{eq:gamma+}
\gamma_{ij}^{+}(\eta,\pmb{k})
& =
-(2\pi)^{3/2} \frac{\sqrt{2k}}{\mpl} 
\left\{ 
e_{ij}^{(R)} e^{-ik\eta} \Big[ \cosh\alpha c_{(R)}(\pmb{k}) + e^{-i\phi} \sinh\alpha c_{(R)}^\dag(-\pmb{k}) \Big]
\right.
\nonumber\\
& \hspace{8em}
\left.
+ e_{ij}^{(L)} e^{ik\eta} \Big[ \cosh\alpha c_{(L)}^\dag(-\pmb{k}) + e^{i\phi} \sinh\alpha c_{(L)}(\pmb{k}) \Big] 
\right\}
\, ,
\\
\label{eq:gamma-}
\gamma_{ij}^{-}(\eta,\pmb{k})
& =
-(2\pi)^{3/2} \frac{\sqrt{2k}}{\mpl} 
\left\{ 
e_{ij}^{(L)} e^{-ik\eta} \Big[ \cosh\alpha c_{(L)}(\pmb{k}) + e^{-i\phi} \sinh\alpha c_{(L)}^\dag(-\pmb{k}) \Big]
\right.
\nonumber\\
& \hspace{8em}
\left.
+ e_{ij}^{(R)} e^{ik\eta} \Big[ \cosh\alpha c_{(R)}^\dag(-\pmb{k}) + e^{i\phi} \sinh\alpha c_{(R)}(\pmb{k}) \Big] 
\right\}
\, .
\end{align}
Thus, we finally find the late-time solution of tensor perturbations, in terms of which we can evaluate the correlation functions, as
\begin{equation}
h_{ij}(\eta\to0,\pmb{k}) = -\frac{H}{k^2} \Big[ \gamma_{ij}^+(0,\pmb{k}) + \gamma_{ij}^-(0,\pmb{k}) \Big] \, ,
\end{equation}
%
It is easy to see that adding together $\gamma_{ij}^++\gamma_{ij}^-$ gives the inverse Fourier transform of \eqref{eq:hij-alpha}. Also note the following property:
\begin{equation}
\gamma_{ij}^\pm(\eta,\pmb{k}) = {\gamma_{ij}^\mp}^\dag(\eta,-\pmb{k}) \, .
\end{equation}
The authors of~\cite{Soda:2011am} argued that this is the key to proving that there is no parity-violating non-Gaussianity in de Sitter space with the BD vacuum, due to the conformal invariance of the correlation functions. In our case, as already known, gravitons are de Sitter invariant in the $\alpha$-vacuum. But, as we will show later, this does not guarantee vanishing parity-violating non-Gaussianity due to the mixing of the positive and negative frequency modes via the Bogoliubov transformation, which boils down to having an imaginary part when evaluating the time integrals in the bispectrum.

\section{Bispectra in $\alpha$-vacuum}
\label{sec:bi}

\subsection{Power spectrum}

Now, we are ready to calculate the three-point correlation functions, or equivalently bispectra, of tensor perturbations. But as a warm-up, let us first compute the power spectrum. Taking the expectation value with respect to the $\alpha$-vacuum of the product of two tensor perturbations at the end of inflation, we find
\begin{align}
\langle h_{ij}(0,\pmb{k})h_{kl}(0,\pmb{q}) \rangle 
& =
\frac{H^2}{k^2q^2} \Big\langle
\big[ \gamma_{ij}^+(0,\pmb{k}) + \gamma_{ij}^-(0,\pmb{k}) \big]
\big[ \gamma_{kl}^+(0,\pmb{q}) + \gamma_{kl}^-(0,\pmb{q}) \big]
\Big\rangle
\nonumber\\
& =
(2\pi)^3 \delta^{(3)}(\pmb{k}+\pmb{q}) \frac{1}{k^3} \frac{H^2}{\mpl^2}
\big[ \cosh(2\alpha) + \cos\phi\sinh(2\alpha) \big] \Big[ \Pi_{ij,kl}^{(L)}(\pmb{k}) + \Pi_{ij,kl}^{(R)}(\pmb{k}) \Big]
\, ,
\end{align}
where we have defined the product of the tensor polarizations as:
\begin{equation}
\Pi_{ij,kl}^{(s)}(\pmb{k}) \equiv e_{ij}^{(s)}(\pmb{k}){e_{kl}^{(s)}}^*(\pmb{k}) \, .
\end{equation}
This result is in agreement with~\cite{Kanno:2022mkx}. We can obtain the contributions from the correlations of the left- and right-circular polarization modes, by making use of \eqref{eq:2pol}, to find
\begin{equation}
h^{(s)} = h_{ij}{e_{ij}^{(s)}}^* \, .
\end{equation}
Then, we can immediately find the tensor power spectrum as
\begin{equation}
\label{eq:powerspectrum}
P_T(k) = \frac{8}{\pi^2} \bigg( \frac{H}{\mpl} \bigg)^2 \big[ \cosh(2\alpha) + \cos\phi\sinh(2\alpha) \big] \, .
\end{equation}
It is now clear that the spectrum for the BD vacuum corresponds to $\alpha=0$.

\subsection{Bispectra}

Now, we compute the bispectra for tensor perturbations in de Sitter background from the interaction Hamiltonian \eqref{eq:H3}. First, we need the following correlation functions:
\begin{align}
\label{eq:-'+}
\langle {\gamma_{ij}^-}'(\eta,\pmb{k}) \gamma^+_{kl}(0,\pmb{q}) \rangle 
& = 
(2\pi)^3 \delta^{(3)}(\pmb{k}+\pmb{q}) \frac{2ik^2}{\mpl^2}
\Big[ e^{ik\eta} \sinh^2\alpha \, \Pi_{ij,kl}^{(R)}(\pmb{k}) - e^{-ik\eta} \cosh^2\alpha \, \Pi_{ij,kl}^{(L)}(\pmb{k}) \Big] 
\, ,
\\ 
\label{eq:+-'}
\langle \gamma^+_{kl}(0,\pmb{q}) {\gamma_{ij}^-}'(\eta,\pmb{k}) \rangle
& = 
(2\pi)^3 \delta^{(3)}(\pmb{k}+\pmb{q}) \frac{2ik^2}{\mpl^2}
\Big[ e^{ik\eta} \cosh^2\alpha \, \Pi_{kl,ij}^{(R)}(\pmb{k}) - e^{-ik\eta} \sinh^2\alpha \, \Pi_{kl,ij}^{(L)}(\pmb{k}) \Big]
\, ,
\\
\label{eq:+'+}
\langle {\gamma_{ij}^+}'(\eta,\pmb{k}) \gamma_{kl}^+(0,\pmb{q}) \rangle 
& =
\langle  \gamma_{kl}^+(0,\pmb{q}) {\gamma_{ij}^+}'(\eta,\pmb{k}) \rangle  
\nonumber\\
& = 
(2\pi)^3 \delta^{(3)}(\pmb{k}+\pmb{q}) \frac{2ik^2}{\mpl^2}
\cosh\alpha \, \sinh\alpha  
\Big[ - e^{-ik\eta} e^{-i\phi} \, \Pi_{ij,kl}^{(R)}(\pmb{k}) + e^{ik\eta} e^{i\phi} \, \Pi_{ij,kl}^{(L)}(\pmb{k}) \Big] 
\, .
\end{align} 
The remaining two-point correlation functions, namely 
$\langle {\gamma^+_{ij}}'(\eta,\pmb{k}) \gamma^-_{kl}(0,\pmb{q}) \rangle$,  
$\langle \gamma^-_{kl}(0,\pmb{q}) {\gamma^+_{ij}}'(\eta,\pmb{k}) \rangle$ and  
$\langle {\gamma^-_{ij}}'(\eta,\pmb{k}) \gamma^-_{kl}(0,\pmb{q}) \rangle 
= \langle \gamma^-_{kl}(0,\pmb{q}) {\gamma^-_{ij}}'(\eta,\pmb{k}) \rangle$, 
have the same expressions as \eqref{eq:-'+}, \eqref{eq:+-'} and \eqref{eq:+'+} respectively, but with interchanged polarizations. To make the results more concise, we make the following definitions for the sum of momenta:
\begin{equation}
\begin{split}
K_{+++} & \equiv k_1 + k_2 + k_3 \, ,
\\
K_{++-} & \equiv k_1 + k_2 - k_3 \, ,
\\
K_{+-+} & \equiv k_1 - k_2 + k_3 \, ,
\\
K_{-++} & \equiv -k_1 + k_2 + k_3 \, ,
\label{eq:Kmpp}
\end{split}
\end{equation}
and put the product of the polarization tensors into a more condensed form:
\begin{equation}
\Pi_{i_1j_1,i_2j_2,i_3j_3}^{(s_1s_2s_3)}(\pmb{k}_1,\pmb{k}_2,\pmb{k}_3)
\equiv
\Pi_{i_1j_1,kl}^{(s_1)}(\pmb{k}_1) \Pi_{i_2j_2,lm}^{(s_2)}(\pmb{k}_2) \Pi_{i_3j_3,mk}^{(s_3)}(\pmb{k}_3)
\, .
\end{equation}
Finally, to perform the integrals, we use the following formula:
\begin{equation}
\int_{-\infty}^0 d\eta \, \eta^5 e^{\pm ik\eta} = \frac{5!}{k^6} \, .
\end{equation}
Given all these, it is straightforward to compute the bispectra by employing the in-in formalism~\cite{Weinberg:2005vy}. We find the following contributions to the bispectra:
\begin{align}
\label{eq:+++}
&
\Big\langle \gamma^+_{i_1j_1}(0,\pmb{k}_1) \gamma^+_{i_2j_2}(0,\pmb{k}_2) \gamma^+_{i_3j_3}(0,\pmb{k}_3) \Big\rangle 
= -\Big\langle \gamma^-_{i_1j_1}(0,\pmb{k}_1) \gamma^-_{i_2j_2}(0,\pmb{k}_2) \gamma^-_{i_3j_3}(0,\pmb{k}_3) \Big\rangle 
\nonumber\\
& = 
(2\pi)^3 \delta^{(3)}(\pmb{k}_1 + \pmb{k}_2 + \pmb{k}_3)
\, 8 \times 5! ib H^5 \frac{(k_1k_2k_3)^2}{\mpl^8}  
\nonumber\\
& 
\quad
\times 
\Bigg\{ [3 \cosh (4 \alpha )+5] \frac{1}{K_{+++}^6}
\Big[ \Pi^{(RRR)}_{i_1 j_1,i_2 j_2, i_3 j_3} (\pmb{k}_1,\pmb{k}_2,\pmb{k}_3) 
+ \Pi^{(LLL)}_{i_1 j_1,i_2 j_2, i_3 j_3} (\pmb{k}_1,\pmb{k}_2,\pmb{k}_3) \Big] 
\nonumber\\
& 
\qquad~
- 2 \sinh^2(2 \alpha) \bigg(
\frac{1}{K_{++-}^6} \Big[ \Pi^{(RRL)}_{i_1 j_1,i_2 j_2, i_3 j_3} (\pmb{k}_1,\pmb{k}_2,\pmb{k}_3)
+ \Pi^{(LLR)}_{i_1 j_1,i_2 j_2, i_3 j_3} (\pmb{k}_1,\pmb{k}_2,\pmb{k}_3) \Big]
\nonumber\\
&
\hspace{8.5em}
+ \frac{1}{K_{+-+}^6}  \Big[ \Pi^{(RLR)}_{i_1 j_1,i_2 j_2, i_3 j_3} (\pmb{k}_1,\pmb{k}_2,\pmb{k}_3)
+ \Pi^{(LRL)}_{i_1 j_1,i_2 j_2, i_3 j_3} (\pmb{k}_1,\pmb{k}_2,\pmb{k}_3) \Big]
\nonumber\\
&
\hspace{8.5em}
+ \frac{1}{K_{-++}^6}  \Big[ \Pi^{(LRR)}_{i_1 j_1,i_2 j_2, i_3 j_3} (\pmb{k}_1,\pmb{k}_2,\pmb{k}_3)
+ \Pi^{(RLL)}_{i_1 j_1,i_2 j_2, i_3 j_3} (\pmb{k}_1,\pmb{k}_2,\pmb{k}_3) \Big] \bigg)
+ \text{5 perms} \Bigg\}
\, ,
\\
\label{eq:++-}
& 
\Big\langle \gamma^+_{i_1j_1}(0,\pmb{k}_1) \gamma^+_{i_2j_2}(0,\pmb{k}_2) \gamma^-_{i_3j_3}(0,\pmb{k}_3) \Big\rangle
\nonumber\\
& = 
(2\pi)^3 \delta^{(3)}(\pmb{k}_1 + \pmb{k}_2 + \pmb{k}_3) 
\, 16 \times 5! i b H^5 \frac{(k_1k_2k_3)^2}{\mpl^8} \sinh(2\alpha)
\nonumber\\
& 
\quad
\times
\Bigg\{ \frac{1}{K_{+++}^6} \Big[ 
A(\alpha,\phi) \, \Pi^{(RRR)}_{i_1 j_1,i_2 j_2, i_3 j_3}(\pmb{k}_1,\pmb{k}_2,\pmb{k}_3) 
+ A^*(\alpha,\phi) \, \Pi^{(LLL)}_{i_1 j_1,i_2 j_2, i_3 j_3} (\pmb{k}_1,\pmb{k}_2,\pmb{k}_3) 
\Big]
\nonumber\\
&
\qquad
- \frac{1}{K_{++-}^6} \Big[ 
B(\alpha,\phi) \, \Pi^{(RRL)}_{i_1 j_1,i_2 j_2, i_3 j_3}(\pmb{k}_1,\pmb{k}_2,\pmb{k}_3) 
+ B^*(\alpha,\phi) \, \Pi^{(LLR)}_{i_1 j_1,i_2 j_2, i_3 j_3} (\pmb{k}_1,\pmb{k}_2,\pmb{k}_3) 
\Big]
\nonumber\\
&
\qquad
+ \frac{1}{K_{+-+}^6} \Big[ \Pi^{(RLR)}_{i_1 j_1,i_2 j_2, i_3 j_3} (\pmb{k}_1,\pmb{k}_2,\pmb{k}_3) 
+ \Pi^{(LRL)}_{i_1 j_1,i_2 j_2, i_3 j_3} (\pmb{k}_1,\pmb{k}_2,\pmb{k}_3) \Big]
\nonumber\\
&
\qquad
+ \frac{1}{K_{-++}^6} \Big[ \Pi^{(LRR)}_{i_1 j_1,i_2 j_2, i_3 j_3} (\pmb{k}_1,\pmb{k}_2,\pmb{k}_3) 
+ \Pi^{(RLL)}_{i_1 j_1,i_2 j_2, i_3 j_3} (\pmb{k}_1,\pmb{k}_2,\pmb{k}_3) \Big]
+ \text{5 perms} \Bigg\}
\, ,
\\
\label{eq:--+}
& 
\Big\langle \gamma^-_{i_1j_1}(0,\pmb{k}_1) \gamma^-_{i_2j_2}(0,\pmb{k}_2) \gamma^+_{i_3j_3}(0,\pmb{k}_3) \Big\rangle
\nonumber\\
& = 
-(2\pi)^3 \delta^{(3)}(\pmb{k}_1 + \pmb{k}_2 + \pmb{k}_3) 
\, 16 \times 5! i b H^5 \frac{(k_1k_2k_3)^2}{\mpl^8} \sinh(2\alpha)
\nonumber\\
& 
\quad
\times
\Bigg\{ \frac{1}{K_{+++}^6} \Big[ 
A^*(\alpha,\phi) \, \Pi^{(RRR)}_{i_1 j_1,i_2 j_2, i_3 j_3}(\pmb{k}_1,\pmb{k}_2,\pmb{k}_3) 
+ A(\alpha,\phi) \, \Pi^{(LLL)}_{i_1 j_1,i_2 j_2, i_3 j_3} (\pmb{k}_1,\pmb{k}_2,\pmb{k}_3) 
\Big]
\nonumber\\
&
\qquad
- \frac{1}{K_{++-}^6} \Big[ 
B^*(\alpha,\phi) \, \Pi^{(RRL)}_{i_1 j_1,i_2 j_2, i_3 j_3}(\pmb{k}_1,\pmb{k}_2,\pmb{k}_3) 
+ B(\alpha,\phi) \, \Pi^{(LLR)}_{i_1 j_1,i_2 j_2, i_3 j_3} (\pmb{k}_1,\pmb{k}_2,\pmb{k}_3) 
\Big]
\nonumber\\
&
\qquad
+ \frac{1}{K_{+-+}^6} \Big[ \Pi^{(RLR)}_{i_1 j_1,i_2 j_2, i_3 j_3} (\pmb{k}_1,\pmb{k}_2,\pmb{k}_3) 
+ \Pi^{(LRL)}_{i_1 j_1,i_2 j_2, i_3 j_3} (\pmb{k}_1,\pmb{k}_2,\pmb{k}_3) \Big]
\nonumber\\
&
\qquad
+ \frac{1}{K_{-++}^6} \Big[ \Pi^{(LRR)}_{i_1 j_1,i_2 j_2, i_3 j_3} (\pmb{k}_1,\pmb{k}_2,\pmb{k}_3) 
+ \Pi^{(RLL)}_{i_1 j_1,i_2 j_2, i_3 j_3} (\pmb{k}_1,\pmb{k}_2,\pmb{k}_3) \Big]
+ \text{5 perms} \Bigg\}
\, ,
\end{align}
where, for convenience, we have evaluated the delta functions as these will be going inside the integrals. We also have defined in \eqref{eq:++-} and \eqref{eq:--+} the following functions of the squeezing parameter $\alpha$ and phase $\phi$:
\begin{align}
A(\alpha,\phi) & \equiv e^{i\phi} \cosh(2\alpha) - e^{-2i\phi} \, ,
\\
B(\alpha,\phi) & \equiv e^{-i\phi} \cosh(2\alpha) + e^{-2i\phi} \, .
\end{align}
The three-point functions $\langle \gamma^+ \gamma^- \gamma^+\rangle$ and $\langle \gamma^- \gamma^+ \gamma^+\rangle$ are comprised by the same two-point functions as in \eqref{eq:++-}, after permuting the momenta. Similarly, $\langle \gamma^- \gamma^+ \gamma^-\rangle$ and $\langle \gamma^+ \gamma^- \gamma^-\rangle$ are comprised by the same two-point functions as in \eqref{eq:--+}, after permuting the momenta.  Note that the three-point correlation functions with mixed polarizations are found to be vanishing in quasi-de Sitter space with the BD initial condition~\cite{Soda:2011am}. Here, in contrast, we find that they are non-trivial in the $\alpha$-vacuum even in perfect de Sitter background due to the mixing by the Bogoliubov transformation.

Having found the three-point correlation functions of the helicity eigenstates, we can next calculate the physical contributions to the bispectra coming from the following tensor bispecrtrum:
\begin{align}
\label{eq:h-bi}
& 
\Big\langle h^{(s_1)}(0,\pmb{k}_1) h^{(s_2)}(0,\pmb{k}_2) h^{(s_3)}(0,\pmb{k}_3) \Big\rangle
\nonumber\\
& = 
(2\pi)^3 \delta(^{(3)}\pmb{k}_1 + \pmb{k}_2 + \pmb{k}_3)
\, 32 \times 5! ib \frac{-H^8}{\mpl^8} 
{e_{i_1j_1}^{(s_1)}}^*(\pmb{k}_1) {e_{i_2j_2}^{(s_2)}}^*(\pmb{k}_2) {e_{i_3j_3}^{(s_3)}}^*(\pmb{k}_3)
\Phi(\alpha,\phi)
\nonumber\\
& 
\quad
\times
\Bigg\{ \frac{3}{K_{+++}^6} \Big[ \Pi^{(LLL)}_{i_1j_1,i_2j_2,i_3j_3}(\pmb{k}_1,\pmb{k}_2,\pmb{k}_3) 
- \Pi^{(RRR)}_{i_1j_1,i_2j_2,i_3j_3}(\pmb{k}_1,\pmb{k}_2,\pmb{k}_3) \Big]
\nonumber\\
& 
\qquad
+ \frac{1}{K_{++-}^6} \Big[ \Pi^{(RRL)}_{i_1j_1,i_2j_2,i_3j_3}(\pmb{k}_1,\pmb{k}_2,\pmb{k}_3) 
- \Pi^{(LLR)}_{i_1j_1,i_2j_2,i_3j_3}(\pmb{k}_1,\pmb{k}_2,\pmb{k}_3) \Big]
\nonumber\\
& 
\qquad
- \frac{1}{K_{+-+}^6} \Big[ \Pi^{(RLR)}_{i_1j_1,i_2j_2,i_3j_3}(\pmb{k}_1,\pmb{k}_2,\pmb{k}_3) 
- \Pi^{(LRL)}_{i_1j_1,i_2j_2,i_3j_3}(\pmb{k}_1,\pmb{k}_2,\pmb{k}_3) \Big]
\nonumber\\
& 
\qquad
+ \frac{1}{K_{-++}^6} \Big[ \Pi^{(LRR)}_{i_1j_1,i_2j_2,i_3j_3}(\pmb{k}_1,\pmb{k}_2,\pmb{k}_3) 
- \Pi^{(RLL)}_{i_1j_1,i_2j_2,i_3j_3}(\pmb{k}_1,\pmb{k}_2,\pmb{k}_3) \Big] \Bigg\}
\, ,
\end{align}
where we have defined the following function which depends only on the parameters of the $\alpha$-vacuum:
\begin{equation}
\label{eq:amp-fct}
\Phi(\alpha,\phi) \equiv 
2 \sinh(2\alpha) \sin\phi \big[ \cosh(2\alpha) + \cos\phi \sinh(2\alpha) \big] \, .
\end{equation}
Note that for the BD vacuum where $\alpha=0$, simply $\Phi(\alpha=0,\phi) = 0$ and \eqref{eq:h-bi} vanishes for any combinations of polarizations. Next, we define the following polarization functions:
\begin{equation}
\label{eq:shape-fct}
F^{(s_1 s_2 s_3)}(k_1,k_2,k_3) 
= {e_{ij}^{(s_1)}}^*(\pmb{k}_1) {e_{jk}^{(s_2)}}^*(\pmb{k}_2) {e_{ki}^{(s_3)}}^*(\pmb{k}_3)
\, ,
\end{equation} 
which, for definite combinations of polarizations, is given by
\begin{align}
\label{eq:pol1}
F^{(RRR)} (k_1,k_2,k_3) 
= F^{(LLL)}(k_1,k_2,k_3) 
& = 
-\frac{K_{+++}^3K_{++-}K_{+-+}K_{--+}}{64k_1^2k_2^2k_3^2} \, ,
\\
\label{eq:pol2}
F^{(RRL)} (k_1,k_2,k_3) 
= F^{(LLR)}(k_1,k_2,k_3) 
& =
-\frac{K_{+++}K_{++-}^3K_{+-+}K_{-++}}{64k_1^2k_2^2k_3^2} \, ,
\\
\label{eq:pol3}
F^{(RLR)} (k_1,k_2,k_3) 
= F^{(LRL)}(k_1,k_2,k_3) 
& =
-\frac{K_{+++}K_{++-}K_{+-+}^3K_{-++}}{64k_1^2k_2^2k_3^2} \, ,
\\
\label{eq:pol4}
F^{(LRR)} (k_1,k_2,k_3) 
= F^{(RLL)}(k_1,k_2,k_3) 
& =
-\frac{K_{+++}K_{++-}K_{+-+}K_{-++}^3}{64k_1^2k_2^2k_3^2} \, .
\end{align}
We need to address each possible combination of the polarizations separately. For the bispectra of purely left- and right-circular polarizations, we have  
\begin{align}
\label{eq:hRRR}
&
\Big\langle h^{(R)}(0,\pmb{k}_1) h^{(R)}(0,\pmb{k}_2) h^{(R)}(0,\pmb{k}_3) \Big\rangle  
= 
- \Big\langle h^{(L)}(0,\pmb{k}_1) h^{(L)}(0,\pmb{k}_2) h^{(L)}(0,\pmb{k}_3) \Big\rangle  
\nonumber\\
& =   
(2\pi)^3 \delta^{(3)}(\pmb{k}_1 + \pmb{k}_2 + \pmb{k}_3)
64 \times 180\, b \frac{H^8}{\mpl^8} \Phi(\alpha,\phi) \frac{F^{(RRR)}(k_1,k_2,k_3)}{K_{+++}^6}
\, .
\end{align}
It is straightforward to see that due to parity violation, there is a sign difference between $\big\langle \big( h^{(L)} \big)^3 \big\rangle$ and $\big\langle \big( h^{(R)} \big)^3 \big\rangle$, so that their difference is not vanishing, i.e. $\big\langle \big( h^{(L)} \big)^3 \big\rangle - \big\langle \big( h^{(R)} \big)^3 \big\rangle \neq 0$. This confirms that there is parity-violating non-Gaussianity in perfect de Sitter background, assuming non-BD initial conditions. Furthermore, we find additional contributions to the bispectrum coming from the mixed polarizations:
\begin{align}
\label{eq:hRRL}
&
\Big\langle h^{(R)}(0,\pmb{k}_1) h^{(R)}(0,\pmb{k}_2) h^{(L)}(0,\pmb{k}_3) \Big\rangle
=  - \Big\langle h^{(L)}(0,\pmb{k}_1) h^{(L)}(0,\pmb{k}_2) h^{(R)}(0,\pmb{k}_3) \Big\rangle
\nonumber\\
& =
(2\pi)^3 \delta^{(3)}(\pmb{k}_1 + \pmb{k}_2 + \pmb{k}_3)
64 \times 60\,b \frac{H^8}{\mpl^8} \Phi(\alpha,\phi) \frac{F^{(RRL)}(k_1,k_2,k_3)}{K_{++-}^6}
\, ,  
\\
\label{eq:hRLR}
&
\Big\langle h^{(R)}(0,\pmb{k}_1) h^{(L)}(0,\pmb{k}_2) h^{(R)}(0,\pmb{k}_3) \Big\rangle
=  - \Big\langle h^{(L)}(0,\pmb{k}_1) h^{(R)}(0,\pmb{k}_2) h^{(L)}(0,\pmb{k}_3) \Big\rangle
\nonumber\\
& =
(2\pi)^3 \delta^{(3)}(\pmb{k}_1 + \pmb{k}_2 + \pmb{k}_3)
64 \times 60\,b \frac{H^8}{\mpl^8} \Phi(\alpha,\phi) \frac{F^{(RLR)}(k_1,k_2,k_3)}{K_{+-+}^6}
\, ,  
\\
\label{eq:hLRR}
&
\Big\langle h^{(L)}(0,\pmb{k}_1) h^{(R)}(0,\pmb{k}_2) h^{(R)}(0,\pmb{k}_3) \Big\rangle
=  - \Big\langle h^{(R)}(0,\pmb{k}_1) h^{(L)}(0,\pmb{k}_2) h^{(L)}(0,\pmb{k}_3) \Big\rangle
\nonumber\\
& =
(2\pi)^3 \delta^{(3)}(\pmb{k}_1 + \pmb{k}_2 + \pmb{k}_3)
64 \times 60\,b \frac{H^8}{\mpl^8} \Phi(\alpha,\phi) \frac{F^{(LRR)}(k_1,k_2,k_3)}{K_{-++}^6}
\, .  
\end{align}
As we will shortly see, these contributions give rise to new shapes for parity-violating non-Gaussianities.

At this point, a clarification is in order. The three-point functions for the mixed polarizations are cyclic. For example, $\langle h^{(L)}(\pmb{k}_1) h^{(R)}(\pmb{k}_2) h^{(R)}(\pmb{k}_3) \rangle$ can be obtained from $\langle h^{(R)}(\pmb{k}_1) h^{(L)}(\pmb{k}_2) h^{(R)}(\pmb{k}_3) \rangle$ by exchanging $\pmb{k}_1$ and $\pmb{k}_2$. Indeed, the labeling of indices is just a matter of convention -- that is, there should be no distinction in the ordering of the momenta. Here for clarity, we keep the order of the momenta as standard, i.e. $(\pmb{k}_1, \pmb{k}_2, \pmb{k}_3)$ and assign shapes to different 
combinations of the polarizations.

\subsection{Shapes of bispectra}

\begin{figure}
\centering
\includegraphics[width=0.5\textwidth]{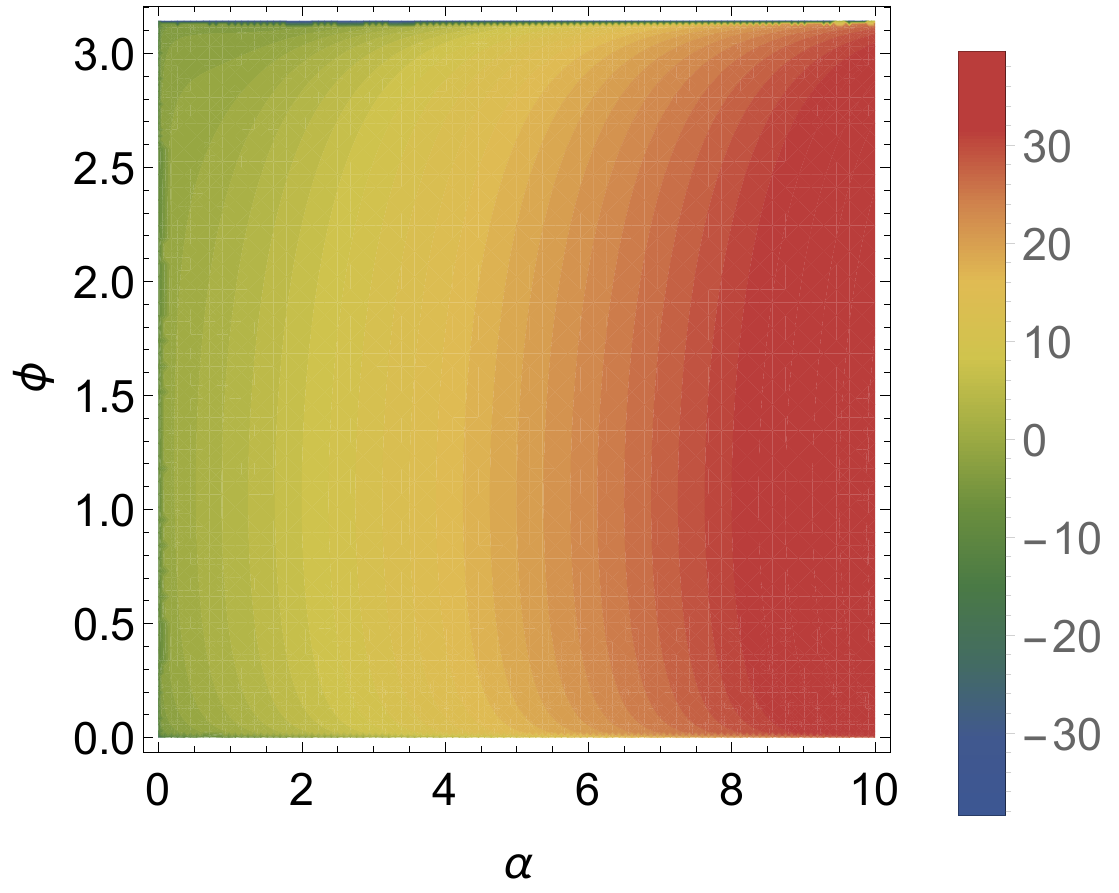}
\caption{Plot of the amplitude function $\log\Phi(\alpha,\phi)$ with $0\leq\alpha\leq10$ and $0\leq\phi\leq\pi$.}
\label{fig:amplitude}
\end{figure}

From \eqref{eq:hRRR}, \eqref{eq:hRRL}, \eqref{eq:hRLR} and \eqref{eq:hLRR}, we can read off immediately the bispectrum of tensor perturbations as 
\begin{equation}
\Big\langle h^{(s_1)}(0,\pmb{k}_1) h^{(s_2)}(0,\pmb{k}_2) h^{(s_3)}(0,\pmb{k}_3) \Big\rangle
\equiv
(2\pi)^3 \delta^{(3)}(\pmb{k}_1 + \pmb{k}_2 + \pmb{k}_3) B^{(s_1s_2s_3)}(k_1,k_2,k_3)
\, .
\end{equation}
As we can see, the bispectrum can be expressed as the product of two contributions: \eqref{eq:amp-fct} and \eqref{eq:shape-fct}, multiplied by the factor $1/K_{\pm\pm\pm}^6$. The former, $\Phi(\alpha,\phi)$, is independent of the triangular configuration of momenta and just rescales, for given values of $\alpha$ and $\phi$, the bispectrum whose shape is determined by $F^{(s_1s_2s_3)}(k_1,k_2,k_3)/K_{\pm\pm\pm}^6$. For this reason, we may call $\Phi(\alpha,\phi)$ as the ``amplitude'' function and $F^{(s_1s_2s_3)}(k_1,k_2,k_3)/K_{\pm\pm\pm}^6$ as the ``shape'' function. 
We show the logarithm of the amplitude function $\Phi(\alpha,\phi)$ in Figure~\ref{fig:amplitude}. Since $\Phi(\alpha,2\pi-\phi) = -\Phi(\alpha,\phi)$, we only present $0\leq\phi\leq\pi$. As we can see, the amplitude function depends exponentially on the squeezing parameter $\alpha$, while the phase $\phi$ does not play a significant role unless $\phi\to0$ or $\phi\to\pi$.

\begin{figure}
	\begin{subfigure}[b]{0.45\textwidth}
		\centering
		\includegraphics[width=\linewidth]{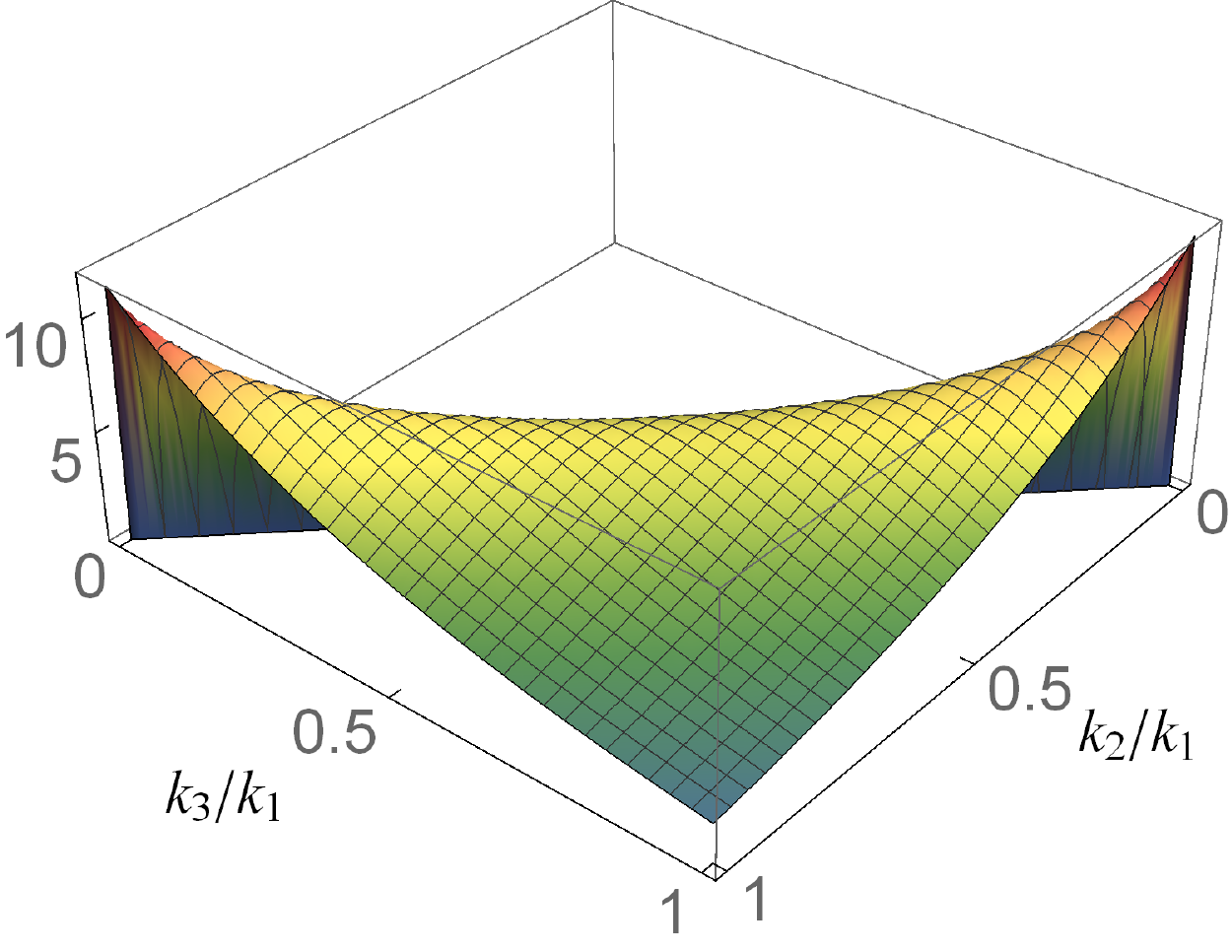}
		\caption{$3\times10^3F^{(RRR)}/K_{+++}^6$}
		\label{fig:ca}
	\end{subfigure}\hfill
	\begin{subfigure}[b]{0.45\textwidth}
		\centering
		\includegraphics[width=\linewidth]{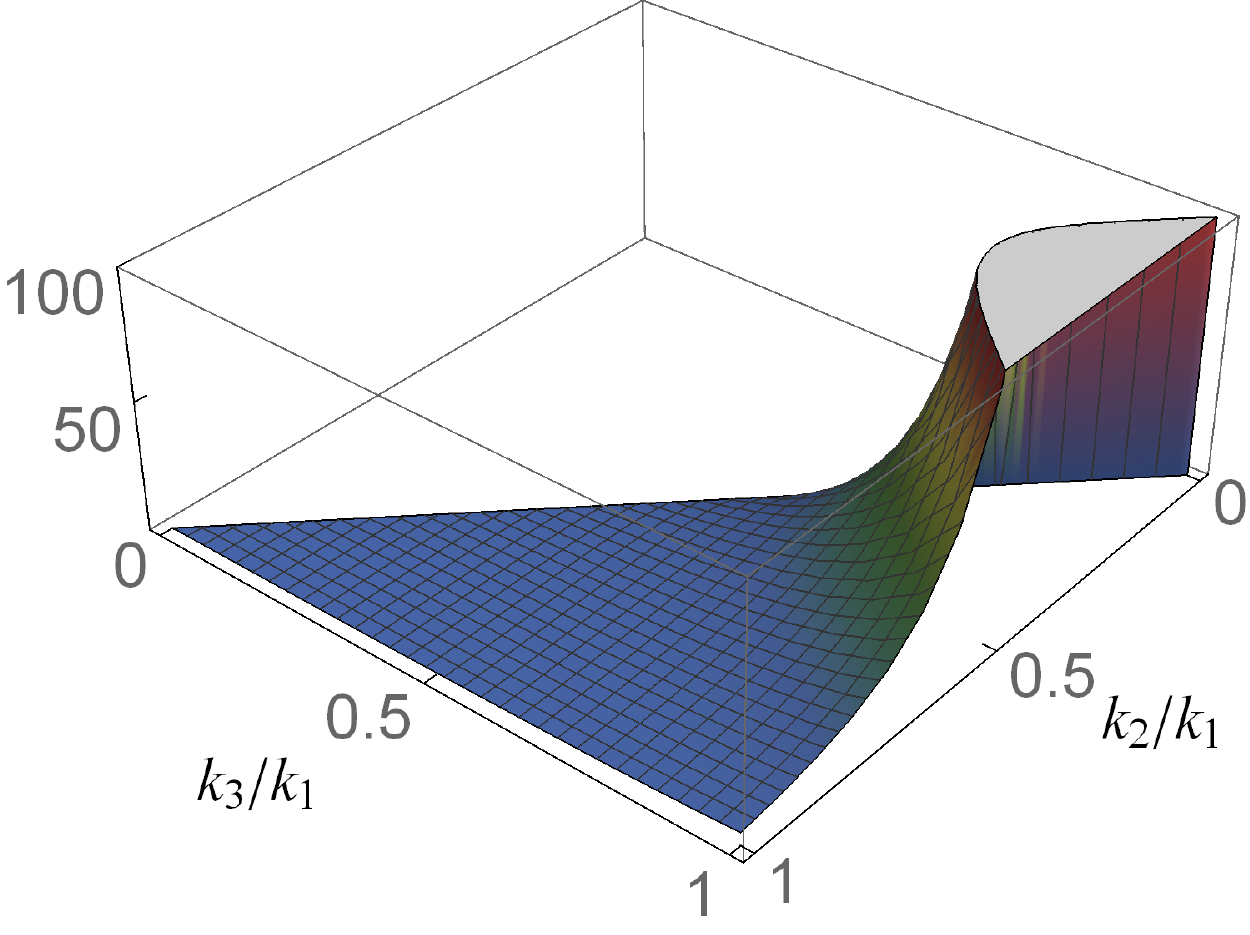}
		\caption{$10^2F^{(RRL)}/K_{++-}^6$}
		\label{fig:cb}
	\end{subfigure}
	\begin{subfigure}[b]{0.45\textwidth}
		\centering
		\includegraphics[width=\linewidth]{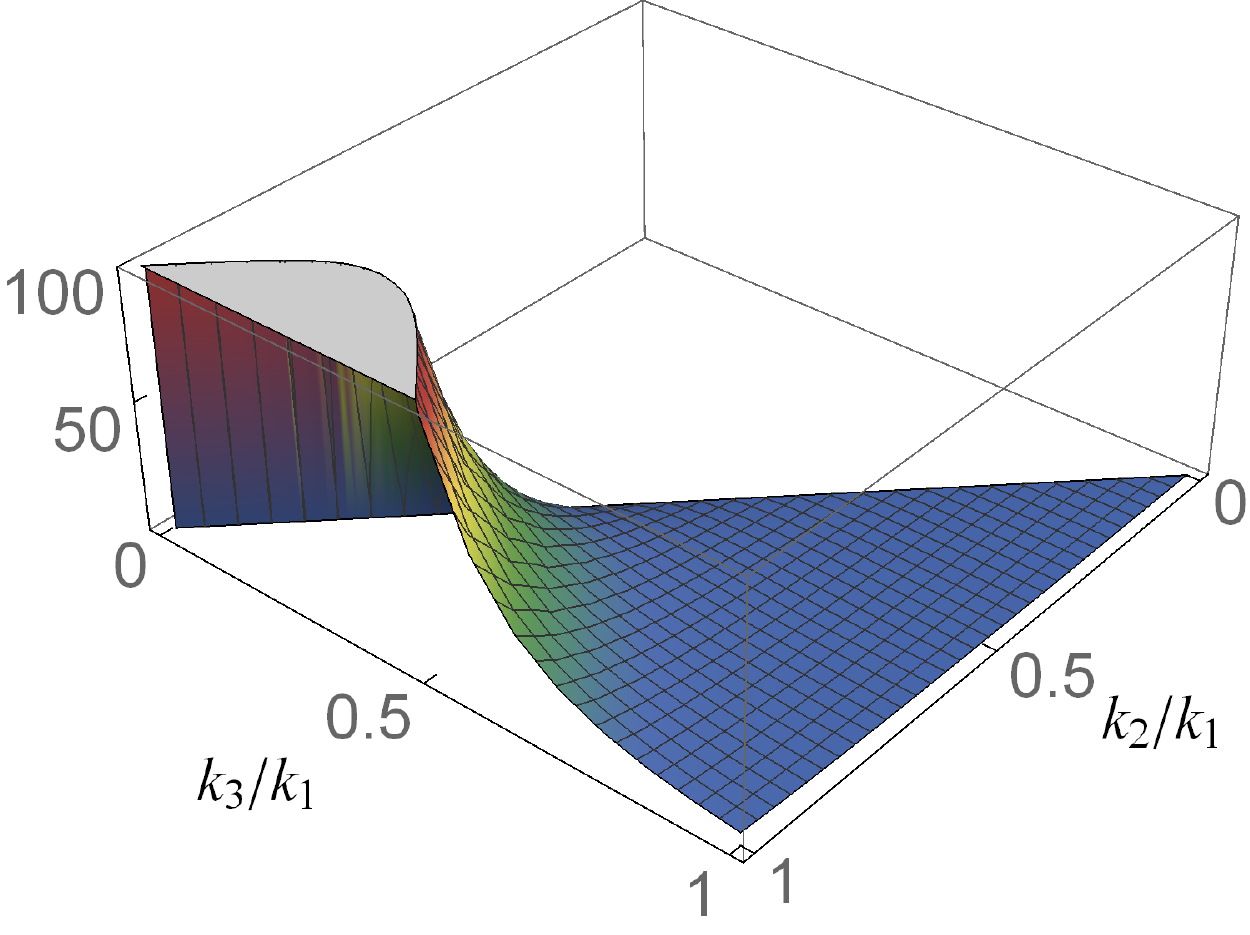}
		\caption{$10^2F^{(RLR)}/K_{+-+}^6$}
		\label{fig:cc}
	\end{subfigure}\hfill
	\begin{subfigure}[b]{0.45\textwidth}
		\centering
		\includegraphics[width=\linewidth]{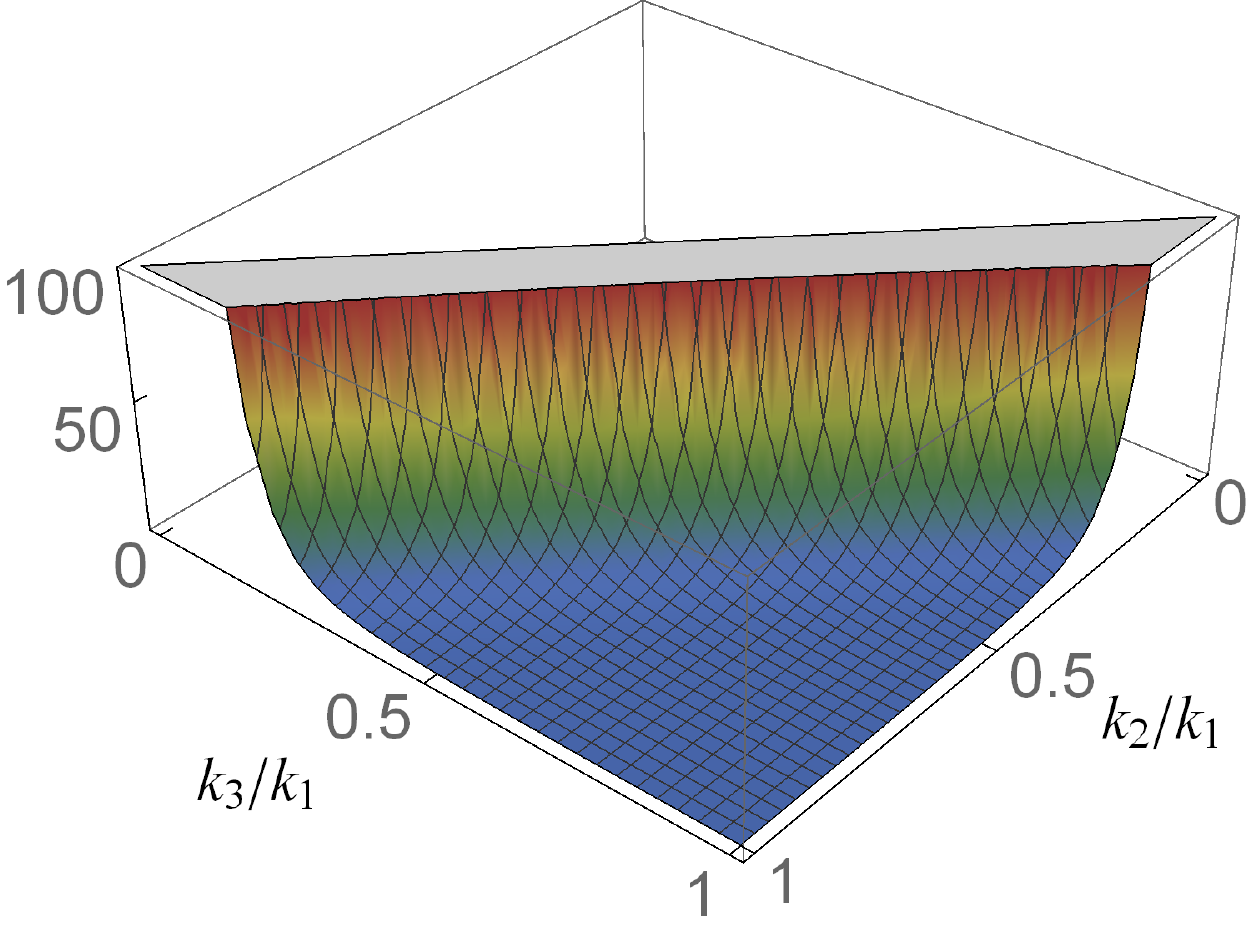}
		\caption{$F^{(LRR)}/K_{-++}^6$}
		\label{fig:1f}
	\end{subfigure}
	\caption{Shapes of the bispectra with possible mixed polarizations.}
\label{fig:shapes}
\end{figure}

We plot the shape function $F^{(s_1s_2s_3)}(k_1,k_2,k_3)$ along with the factor $1/K_{\pm\pm\pm}^6$ in Figure~\ref{fig:shapes}. The contributions from the purely left- and right-polarizations are subdominant. Meanwhile, the contributions from the mixed polarizations peak in the squeezed limit, while the dominant contribution comes from the flattened configuration, i.e. $k_2 \approx k_3 \to k_1/2$. 
The fact that our configurations peak in the flattened and squeezed configurations is not surprising, as this is customary in theories of inflation with excited initial conditions~\cite{Chen:2006nt,Holman:2007na,Meerburg:2009ys,Mylova:2021eld}. Indeed, these shapes were previously found in standard Einstein gravity with the $\alpha$-vacuum~\cite{Kanno:2022mkx}. But it is worth to emphasize that as far as we know, this is the first time in the literature that we obtain a flattened configuration for parity-violating theories of gravity. Interestingly, a flattened shape for tensor non-Gaussianity can be relevant for direct gravitational wave experiments, such as interferometers and pulsar timing arrays. In fact, while Shapiro time-delay effects~\cite{DeLuca:2019qsy} tend to decorrelate non-Gaussian gravitational wave signals, tensor non-Gaussianities enhanced in flattened~\cite{Tasinato:2022xyq} and squeezed~\cite{Dimastrogiovanni:2019bfl} configurations  can be detectable.

\begin{figure}
\centering
\includegraphics[width=0.5\textwidth]{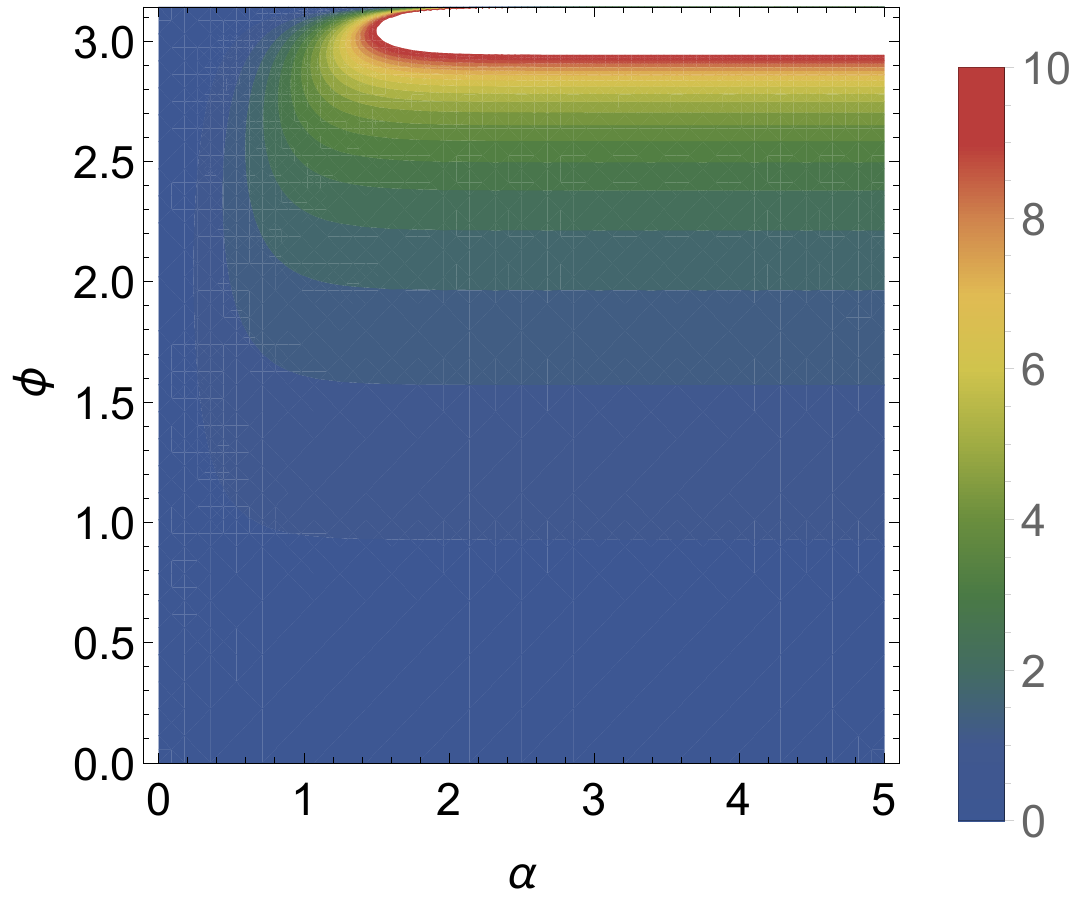}
\caption{Plot of the non-linear parameter $\fnl$ given by \eqref{eq:fnl-amp} with $0\leq\alpha\leq5$ and $0\leq\phi\leq\pi$.}
\label{fig:fnl}
\end{figure}

Next, we focus on the non-linear parameter $\fnl$. On general ground, extending the standard definition of $\fnl$ for local-type non-Gaussianity~\cite{Komatsu:2001rj}, we can define the polarization-dependent $\fnl$ as the bispectrum divided by the power spectrum squared:
\begin{equation}
\label{eq:fnl}
\fnl^{(s_1s_2s_3)}(k_1,k_2,k_3) 
\equiv 
\frac{B^{(s_1s_2s_3)}(k_1,k_2,k_3)}{P_T^2} 
\, ,
\end{equation} 
where the power spectrum $P_T$ is given by \eqref{eq:powerspectrum}. Focusing on the amplitude of $\fnl$ rather than the detailed shape dependence, we can approximate $\fnl$ as
\begin{equation}
\label{eq:fnl-amp}
\fnl \approx 
\frac{\sinh(2 \alpha)  \sin \phi}{\cosh(2\alpha) + \cos \phi \sinh(2\alpha)}
\, .
\end{equation} 
In Figure~\ref{fig:fnl}, we present \eqref{eq:fnl-amp} as a function of $\alpha$ and $\phi$. In particular, as $\phi$ approaches $\pi$, $\fnl$ is exponentially enhanced as
\begin{equation}
|\fnl| \approx e^{2\alpha} \, \big( e^{2\alpha}|\phi-\pi| \big) \, ,
\end{equation} 
for $e^{2\alpha}|\phi-\pi|\ll 1$ and $e^{2\alpha}\gg1$.
This dominates for a large squeezing parameter $\alpha$, assuming a fairly large deviation from the BD vacuum, which could be easily realized if the energy scale of inflation is very low, say a few GeV: See discussions in~\cite{Kanno:2022mkx} and references therein. Note that this is a relative enhancement, because the power spectrum is exponentially suppressed for $\phi \sim \pi$. On the other hand, if $\phi=\pi$ or $\alpha=0$,  we recover the BD results with $\fnl=0$.

\subsection{Helicity conservation and flattened shape}

Before we conclude, it is important to discuss the conservation of graviton helicity in the flattened configuration. Using our convention of normalization with respect to $k_1$, as can be seen in Figure~\ref{fig:1f}, this is given by $\langle h^{(L)}(\pmb{k}_1) h^{(R)}(\pmb{k}_2) h^{(R)}(\pmb{k}_3) \rangle$ in \eqref{eq:hLRR}.  In~\cite{Agrawal:2018mrg} it was argued that this graviton vertex is not allowed due to helicity conservation and thus the flattened shape is forbidden. Indeed, from a particle physics point of view, the flattened configuration is described by the forward scattering of two right-handed gravitons $k_2+k_3 =k_1$. In such a system the spin cannot be conserved and therefore the cross section is vanishing. Similarly, due to angular momentum conservation, the shape function $F^{(LRR)}$ vanishes and the flattened bispectrum does not exist. It is claimed in~\cite{Akama:2020jko} that this vanishing of the flattened shape persists beyond Einstein gravity.

In what follows, we argue that this is only true when the physics under consideration is deep inside the horizon where the particle physics picture is valid, and explain  why the flattened shape is non-vanishing in an expanding universe. On general grounds, in the in-in formalism one obtains the expectation value of the product of operators $h^{(L)}h^{(R)}h^{(R)}$ at some arbitrary time $\eta$. Therefore, in order to better understand the behaviour of the three-point function in the flattened configuration, it is best to \textit{first} evaluate the three-point function at some arbitrary time $\eta$ and \textit{then} explore various possibilities by taking appropriate limits. This is the approach we take here. There are several contributions to the three-point function $\langle h^{(L)}h^{(R)}h^{(R)} \rangle$, but it all essentially boils down into evaluating an expression that has, approximately, the following form\footnote{Here, we are ignoring various overall factors that ensure the non-vanishing parity-violating non-Gaussianity in de Sitter space.}:
\begin{equation}
\Big\langle h^{(L)}(\eta,\pmb{k}_1) h^{(R)}(\eta,\pmb{k}_2) h^{(R)}(\eta,\pmb{k}_3) \Big\rangle 
\sim 
H^8 (-k_1 \eta)(-k_2 \eta)(-k_3 \eta) \int^\eta_{\eta_0} d {\eta'} {\eta'}^5 e^{-i K_{-++} \eta^\prime} 
F^{(LRR)}(k_1,k_2,k_3) \, .
\label{eq:flattened-integral}
\end{equation} 
%
%
Here, the factors $(-k_i\eta)$ outside the time integral come from \eqref{eq:helicity-modes}, \eqref{eq:gamma'}, \eqref{eq:gamma+} and \eqref{eq:gamma-} with $a=-1/(H\eta)$ in de Sitter space, evaluated at some arbitrary time $\eta$, as opposed to the usual prescription of taking $\eta \rightarrow 0$. 
Computing this integral simply gives
\begin{equation}
\Big\langle h^{(L)}(\eta,\pmb{k}_1) h^{(R)}(\eta,\pmb{k}_2) h^{(R)}(\eta,\pmb{k}_3) \Big\rangle 
\sim  
H^8 (-k_1 \eta)(-k_2 \eta)(-k_3 \eta) F^{(LRR)}(k_1,k_2,k_3) \big[Y(\eta) - Y(\eta_0)\big] \, ,
\end{equation} 
where $Y(\eta)$ is obtained by performing the time integral in \eqref{eq:flattened-integral}:
\begin{equation}
Y(\eta) 
= \frac{120 e^{-iK_{-++}\eta}}{K_{-++}^6} 
\left[1 + i K_{-++}\eta - \frac{1}{2} ( K_{-++}\eta)^2 - \frac{1}{6} i (  K_{-++}\eta)^3 
+ \frac{1}{24} ( K_{-++}\eta )^4 + \frac{1}{120}i (  K_{-++}\eta)^5 \right] \, .
\label{eq:yeta}
\end{equation} 
The system can be correctly placed in the initial  $\alpha$-vacuum state by redefining the time $\eta$ to have a small imaginary component, i.e. $\eta \rightarrow \eta + i \varepsilon \eta$ with $\varepsilon \ll 1$. This can be done by applying the standard prescription~\cite{Maldacena:2002vr} of adding a small exponential damping term in the exponent for $Y(\eta_0)$ and taking $\eta_0 \rightarrow -\infty$. Then $Y(\eta_0) = 0$ and we obtain the following expression in terms of an arbitrary time $\eta$:
\begin{align}
& 
\frac{15K_{+++}K_{++-}K_{+-+}}{8 k_1^2 k_2^2 k_3^2 K_{-++}^3} e^{-i K_{-++}\eta} 
H^8 (-k_1 \eta)(-k_2 \eta)(-k_3 \eta)
\nonumber\\
& \times \left[1 + i K_{-++}\eta - \frac{1}{2} ( K_{-++}\eta)^2 -\frac{1}{6} i (  K_{-++}\eta)^3 
+ \frac{1}{24} ( K_{-++}\eta )^4 + \frac{1}{120}i (  K_{-++}\eta)^5 \right] \, ,
\label{eq:flattened-integral2}
\end{align} 
where we have substituted for the explicit form of the shape function $F^{(LRR)}(k_1,k_2,k_3)$. It is now clear to see that in \eqref{eq:flattened-integral2} there are cancellations between the scale dependence in $Y(\eta)$ and $F^{(LRR)}$. To understand what this means, we need to examine the three-point function during different stages of its cosmological evolution.

\begin{itemize}

\item 
To see how each term behaves deep within the horizon $-k_i\eta \gg 1$, let us rewrite $\eta$ in terms of $a=-1/(H\eta)$. Then, each power of $(-K_{-++}\eta)$ and $(-k_i\eta)$ gives a factor $H^{-1}$. From this, we can see that the term proportional to $(K_{-++}\eta)^5$ in the extreme sub-horizon limit goes like
\begin{equation}
\frac{k_1k_2k_3}{(aH)^3} H^8 \bigg( \frac{K_{-++}}{aH} \bigg)^5 \, ,
\end{equation}
so that no  $H$ remains and therefore is not affected by curvature effects, which are understood to be of $\mathcal{O} (H^2)$. Now taking the flattened limit $K_{-++}\to0$ this term naturally vanishes. Similarly, the term $(K_{-++}\eta)^4$ is proportional to $\mathcal{O}(H)$ and therefore also vanishes in the flattened limit. 

\item 
The remaining terms in \eqref{eq:flattened-integral2} are affected by curvature effects, being proportional to powers of $\mathcal{O} (H^2)$ or larger. It is now clear that taking the flat space-time limit $H \rightarrow 0$ makes them vanishing well within the sub-horizon regime. This is in agreement with the requirement of helicity conservation as in~\cite{Agrawal:2018mrg} so that, as expected, the particle physics picture persists well within the sub-horizon regime.

\item 
On the other hand, as we move away from the deep sub-horizon limit, the remaining terms, which are at least $\mathcal{O}(H^2)$, are either finite or divergent in the flattened limit. Here, there is no contradiction with angular momentum conservation. Indeed, since the shape function is a combination of the three polarizations at a vertex, which is also present in the flat space-time limit, it reflects the angular momentum conservation and hence vanishes in the flattened limit. The important point here is that once curvature effects are involved, then there are cancellations between $F^{(LRR)}$  and the non-trivial scale dependence in $Y(\eta)$, as seen in \eqref{eq:flattened-integral2}, which contribute to the dominant result in \eqref{eq:hLRR} when $-K_{-++}\eta\to0$.

\end{itemize}

We have no idea about the exact meaning of the divergence in the flattened limit. We speculate that the three-point function in real space would be finite as long as we choose three different points. But this discussion is beyond the scope of this article and deserves more careful considerations in the future, especially given the possibility that we may be able to observe such a signal. There, we know that in practice we can never achieve a \textit{perfect} squeezed or flattened configuration due to the finite resolution of our observational instruments.

\section{Conclusions}
\label{sec:conc}

In this work, we studied parity violation in exact de Sitter space with the $\alpha$-vacuum initial conditions. We find that even a small deviation from the BD vacuum can result in parity-violating non-Gaussianity in gravity. This is in contrast to previous results in the literature, which concluded that there is no parity-violating non-Gaussianity in perfect de Sitter background due to the isometries of de Sitter space. Indeed, it is well known that gravitons in the $\alpha$-vacuum respect the de Sitter symmetries. The difference here is that the Bogoliubov transformation introduces an imaginary part in the bispectrum which is, otherwise, absent in the parity-violating theory with BD initial condition.

We also showed that in this setup, the mixed tensor polarizations can become important, peaking in different momentum configurations, with the dominant contribution coming in the flattened limit. This is a new shape for parity-violating tensor non-Gaussianity that has not been seen before in the literature and it's worth further investigation. Finally, we find that the bispectrum amplitude can be potentially enhanced for large squeezing parameter $\alpha$ when the phase $\phi$ approaches $\pi$. Therefore, our results provide a distinctive observational signature for the parity-violating gravity sector that leads to the exciting possibility to test the initial conditions of inflation with future detections. It would be interesting to study the resulting CMB cross-bispectra of the temperature and polarization modes. We hope to examine this in a future work.

\subsection*{Acknowledgments}

We would like to thank Sugumi Kanno and Vicharit Yingcharoenrat for initial collaboration, and Shingo Akama, Wei-Chen Lin, Marianthi Moschou, Jiro Soda and Gianmassimo Tasinato for advice in relation to this work. 
JG and MM are grateful to the Kavli Institute for the Physics and Mathematics of the Universe for hospitality during the workshop ``Non-linear aspects of cosmological gravitational waves'', where this work was initiated.
JG and MM are supported in parts by the Mid-Career Research Program (2019R1A2C2085023) through the National Research Foundation of Korea Research Grants.
JG also acknowledges the Korea-Japan Basic Scientific Cooperation Program supported by the National Research Foundation of Korea and the Japan Society for the Promotion of Science (2020K2A9A2A08000097) and the Ewha Womans University Research Grant of 2022 (1-2022-0606-001-1).
MS is supported in part by JSPS KAKENHI Grants Nos.~19H01895, 20H04727, and 20H05853.
JG is grateful to the Asia Pacific Center for Theoretical Physics for hospitality while this work was under progress.

\bibliography{bibi} 
\bibliographystyle{utphys}

\end{document}